\documentclass[reqno]{amsart}
\usepackage{amsfonts}
\usepackage{amssymb}
\usepackage{amsbsy}
\newtheorem{theorem}{Theorem}[section]
\newtheorem{proposition}[theorem]{Proposition}
\newtheorem{definition}[theorem]{Definition}
\def\fl{}
\usepackage{eucal}

\newcommand{\SL}{\mathrm{SL}}

\newcommand{\Aut}{\mathrm{Aut}}
\newcommand{\bom}{\boldsymbol{\omega}}
\newcommand{\rmd}{\mathrm{d}}
\newcommand{\rmi}{\mathrm{i}}
\newcommand{\btau}{\boldsymbol{\tau}}

\newcommand{\bth}{\boldsymbol{\theta}{}}

\newcommand{\bOm}{\boldsymbol{\Omega}}

\newcommand{\bTheta}{\boldsymbol{\Theta}}
\newcommand{\bTh}{\bTheta}

\newcommand{\bUp}{\bUpsilon}
\newcommand{\bUpsilon}{\boldsymbol{\Upsilon}}
\newcommand{\bGamma}{\boldsymbol{\Gamma}}
\newcommand{\bOmega}{\boldsymbol{\Omega}}

\newcommand{\be}{{\boldsymbol{e}}}

\newcommand{\bA}{{\boldsymbol{A}}}

\newcommand{\Lin}{\mathsf{Lin}}

\renewcommand{\O}{\mathsf{O}}

\renewcommand{\o}{\mathfrak{o}}
\newcommand{\txi}{\tilde{\xi}}
\newcommand{\tR}{\tilde{R}}

\newcommand{\cR}{\mathcal{R}}
\newcommand{\cI}{\mathcal{I}}

\newcommand{\ttau}{\tilde{\tau}}
\newcommand{\tmu}{\tilde{\mu}}

\newcommand{\tLambda}{\tilde{\Lambda}}

\newcommand{\Ga}{\Gamma}
\newcommand{\bG}{\boldsymbol{\Gamma}}
\newcommand{\tG}{\tilde{\Gamma}}
\newcommand{\tUp}{\tilde{\Upsilon}}
\newcommand{\tXi}{\tilde{\Xi}}

\def\bomu^#1_#2{\bom^{#1}_{\ #2}}
\def\bOmu^#1_#2{\bOm^{#1}_{\ #2}}
\def\Gu^#1_#2{\Gamma^{#1}_{\ #2}}
\def\tGu^#1_#2{\tilde{\Gamma}^{#1}_{\ #2}}
\def\R^#1_#2{R^{#1}_{\ #2}}
\def\M^#1_#2{M^{#1}_{\ #2}}
\def\C^#1_#2{C^{#1}_{\ #2}}
\def\Up^#1_#2{\Upsilon^{#1}_{\ #2}}

\newcommand{\Rset}{\mathbb{R}}

\newcommand{\N}{{\mathsf{N}}}

\newcommand{\so}{\mathfrak{so}}

\newcommand{\fg}{\mathfrak{g}}

\newcommand{\CH}{\mathrm{CH}}

\newcommand{\F}{\mathsf{F}}

\newcommand{\tGamma}{\tilde{\Gamma}}

\newcommand{\hN}{\hat{N}}

\newcommand{\vm}{{\boldsymbol{m}}}
\newcommand{\bvm}{{\bar{\vm}}}
\newcommand{\vell}{\boldsymbol{\ell}}
\newcommand{\vn}{\boldsymbol{n}}

\begin{document}

\markboth{R. Milson, N. Pelavas}{Curvature homogeneity bound}
\title{The curvature homogeneity bound for Lorentzian four-manifolds}
\author{R. Milson}

\address{Dept. Mathematics and Statistics, Dalhousie University\\
  Halifax NS B3H 3J5, Canada } \email{rmilson@dal.ca}
\author{N. Pelavas} 
\email{pelavas@mathstat.dal.ca}
\maketitle

\begin{abstract}
  We prove that a four-dimensional Lorentzian manifold that is
  curvature homogeneous of order $3$, or $\CH_3$ for short, is
  necessarily locally homogeneous. We also exhibit and classify
  four-dimensional Lorentzian, $\CH_2$ manifolds that are not
  homogeneous.  The resulting metrics belong to the class of null
  electromagnetic radiation, type N solutions on an anti-de Sitter
  background. These findings prove that the four-dimensional
  Lorentzian Singer number $k_{1,3}=3$, falsifying some recent
  conjectures\cite{gilkey}.  We also prove that invariant
  classification for these proper $\CH_2$ solutions requires
  $\nabla^{(7)}R$, and that these are the unique metrics requiring the
  seventh order.
\end{abstract}


\section{Introduction}
The invariant classification (IC) of spacetimes is central importance
in general relativity.  Let $(M,g)$ be a pseudo-Riemannian manifold,
and let $R^i=\nabla^{i}R$ denote the $i$th-order covariant derivative
of the Riemann curvature tensor.  Throughout, we assume that the
algebraic type of the curvature tensor and its covariant derivatives
is constant.  This means that the curvature and its covariant
derivatives can be normalized to some standard form so that there is a
well defined automorphism group $G_i$ of $R^i$. We set $N_i = \dim
G_i$.  The general equivalence problem for pseudo-Riemannian geometry
was solved by Elie Cartan, who proved that $R, R^1,\ldots, R^q$, up to
sufficiently high order classifies the metric up to a diffeomorphism
\cite{cartan}.  Let $q_M$ denote the smallest order required for the
invariant classification (IC) of $M$.  Cartan established the bound
$q_M\leq n(n+1)/2$; here $n(n+1)/2$ is the dimension of the
corresponding orthogonal frame bundle.

Motivated by applications to GR, Brans\cite{brans65} and Karlhede
\cite{karlhede80} showed how to recast the equivalence problem in
terms of differential invariants on the base manifold.  One advantage
of Karlhede's algorithm (see below) is that it improves the general IC
bound to $q_M\leq N_0 + n+1$.  The IC algorithm was refined and
implemented in a computer algebra system by MacCallum, \AA man
\cite{MacCAman86}, and others \cite{maccallumskea94}. See
\cite{PSdI2000} for a recent review.  Here is the algorithm, largely
as it appears in \cite[Section 9.2]{ES}.  Let $\eta_{ab}$ be a
constant, non-degenerate quadratic form having the same signature as
the metric $g$.  Henceforth, we use $\eta_{ab}$ to raise and lower
frame indices.  Let $\O(\eta)$ denote the $n(n-1)/2$ dimensional Lie
group of $\eta$-orthogonal transformations, and say that a coframe
$\bth^a$ is $\eta$-orthogonal if
\begin{equation}
  \label{eq:eta-orthog}
  g = \eta_{ab}\bth^a \bth^b
\end{equation}
\noindent\textbf{The  Karlhede IC algorithm}
\begin{itemize}
\item[1.] Set $q=0, G_{-1} = O(\eta), t_{-1} = 0$.  All
  $\eta$-orthogonal frames are permitted.
\item[2.] Compute $R^q$ relative to a permitted $\eta$-orthogonal
  frame.
\item[3.]  Determine $G_q\subset G_{q-1}$, the automorphism group of
  $R^{(q)}:=\{ R^0,R^1,\ldots, R^q \}$.
\item[4.] Restrict the frame freedom to $G_q$ by putting
  $R^q$ into standard form (normalizing some components to a
  constant, for instance.)
\item[5.] Having restricted the frame freedom, the functions in the
  set $R^{(q)}$ are differential invariants.  Let $t_{q}$ be the
  number of independent functions over $M$ in $R^{(q)}$.
\item[6.] If $N_{q} < N_{q-1}$ or $t_{q}>t_{q-1}$, then
  increase $q$ by one, and go to step 2.
\item[7.] Otherwise, the algorithm terminates.  The differential
  invariants in $R^{(q-1)}$ furnish essential coordinates.  The isometry
  group has dimension $n-t_{q-1}+N_{q-1}$. The orbits have dimension
  $n-t_{q-1}$.  The integer $q_M=q$ is the IC order.
\end{itemize}
\noindent
In principle, the invariant classification and the equivalence
problems are solved at step 7 because the essential coordinates
obtained via the algorithm allow the metric to be expressed in a
canonical form that incorporates the other differential invariants as
essential constants and essential functional parameters.

An optimal bound on $q_M$ where $M$ is a Lorentzian, 4-dimensional
manifold is of particular interest in classical general relativity.
The well-known Petrov-Penrose classification of the Weyl tensor gives
$N_0=0$ for Petrov types I, II, III; $N_0\leq 2$ for types N and D;
and $N_0\leq 3$ for type O\footnote{Here, one has to consider the
  possible symmetries of the Ricci tensor.}.  Hence, $q_M\leq 5$ for
types I, II, III; $q_M\leq 7$ for Petrov types N, D; and $q_M\leq 8$
for type O. These bounds have been improved, and it is now known that
$q_M\leq 6$ for a type D spacetime \cite{coldinv93}, and $q_M\leq 6$
for a type O spacetime \cite{PS2000}.

The question of whether the 7th order bound for type N spacetimes was
sharp or whether it could be improved remained open for over 20 years.
Recently, the present authors exhibited a family of type N exact
solutions for null electromagnetic radiation on an anti-de Sitter
background for which $q_M=7$ , and thereby established that Karlhede's
bound of $q_M\leq 7$ was indeed sharp \cite{kbsharp}.  In the present
paper, we give a detailed derivation of the exact solutions in
question, and prove that these metrics are, essentially, the unique
spacetimes for which $q_M=7$.

Our approach is to consider the restricted IC problem for the class of
proper, curvature homogeneous geometries and to express the curvature
homogeneity condition in terms of an appropriate set of field
equations.
\subsection{Curvature homogeneity and invariant classification}
A pseudo-Riemannian manifold is \emph{curvature homogeneous} of order
$k$, or $\CH_k$ for short, if the components of the curvature tensor
and its first $k$ covariant derivatives are constant relative to some
choice of frame.  We say that $M$ is \emph{properly} $\CH_k$ if it
belongs to class $\CH_k$, but does not belong to class $\CH_{k+1}$
\cite{boeckx}.  The $\CH$ class includes all homogeneous geometries,
because a homogeneous space is curvature homogeneous to all orders.
Thus, a (locally) homogeneous manifold is $\CH_k$ for all $k$, but not
properly $\CH_k$ for any $k$.  The following remarkable result was
originally proved by Singer in the Riemannian context\cite{singer} and
extended to arbitrary signatures in \cite{PoSp}.
\begin{theorem}[Singer, Podesta and Spiro]
  For every signature $(a,b)$, there exists an integer $k$, such that
  if $M$ is $\CH_{k}$, then necessarily $M$ is locally homogeneous.
\end{theorem}
\noindent
Following Gilkey\cite{gilkey}, we use $k_{a,b}$ to denote the smallest
such integer $k$.  The proof of the theorem utilizes an integer
invariant $k_M$, defined to be the smallest $k$ such that $N_k =
N_{k+1}$ \cite{BuDj00,boeckx}.  Singer established the following.
\begin{theorem}[Singer's criterion]
  \label{thm:singcrit}
  If $M$ is curvature homogeneous of order $k_M+1$, then $M$
  is locally homogeneous.  
\end{theorem}
\noindent
Consequently, if $M$ is properly $\CH_k$, then necessarily, $k\leq
k_M$.  It follows that $k_{a,b} = \max \{k_M+1\}$ where $M$ runs over
the class of proper curvature-homogeneous manifolds of signature
$(a,b)$.  Also note that Singer's criterion follows as a special case
of the Karlhede algorithm.  Indeed, $M$ is a homogeneous space if and
only if all differential invariants are essential constants (the
structure constants of the corresponding Lie algebra.)  Thus, $M$ is a
homogeneous space if and only if $t_{k_M+1}=0$.  The latter condition
is equivalent to $M$ being curvature homogeneous of order $k_M+1$.

As we will show, the class of proper $\CH$ manifolds plays a key role
in the search for geometries with a maximal $q_M$.  Already in
\cite{coldinv93}, Collins and d'Inverno showed that the conditions for
an IC order of $q_M=7$ are very stringent.  Without naming it as such,
their necessary conditions (shown below) suggest a proper $\CH_2$
geometry.
\begin{itemize}
\item[(C1)]  The components of the curvature tensor must be constants.
\item[(C2)] The invariance group at zeroth order $G_0$, must have dimension 2.
\item[(C3)] The dimension of the invariance group and the number of
  functionally independent components must not both change on
  differentiating.
\item[(C4)] We must produce at most one new functionally independent
  component on differentiating.
\item[(C5)] The dimension of the invariance group must go down by at
  most one dimension on differentiating.
\end{itemize}
These conditions imply that $q_M=7$ can be achieved if
\begin{eqnarray}
  \label{eq:tqcount}
  &&(t_0, t_1, t_2, t_3, t_4, t_5, t_6, t_7) = (0,0,0,1,2,3,4,4);\\
  \label{eq:dimgicount}
  &&(N_0,N_1, N_2,N_3,\ldots ) = (2,1,0,0,\ldots).
\end{eqnarray}
It is conceivable that a $q_M=7$ might be achieved with a different
sequence of $t_i$ and $N_i$, but that would require
$N_i=N_{i+1}>N_{i+2}$ for some $i$.  Such a phenomenon is called
pseudo-stabilization and it is known to be highly
atypical\cite{olver}.  Since $N_{k_M} = N_{k_M+1}$, we can also say
that the curvature automorphism groups pseudo-stabilize if there
exists an $i>k_M+1$ such that $N_i< N_{k_M}$.  If we exclude the
possibility of pseudo-stabilization then the Collins--D'Inverno
conditions describe a proper, curvature homogeneous geometry.
\begin{proposition}
  Suppose that there exists an $M$ such that $q_M = N_0 + n+1$, i.e.
  the Karlhede bound is sharp.  Also, suppose that the curvature
  automorphism groups do not pseudo-stabilize. Then, $M$ is properly
  $\CH_k$ where $k={N_0}$.
\end{proposition}
In the case of type N spacetimes, if the Karlhede bound $q_M\leq 7$
really were sharp, and if we exclude the possibility of
pseudo-stabilization, then we are forced to consider the existence of
a proper $\CH_2$ geometry.

To put it another way, the value of Gilkey's integer $k_{1,3}$ is
crucial, because if $k_{1,3}\leq2$, then a proper $\CH_2$ Lorentzian
manifold does not exist.  Let us review what is known about bounds on
$k_M$.  In his original paper \cite{singer} Singer established the
bound $k_M < n(n-1)/2$; here the right-hand side is the dimension of
the orthogonal group. In the Riemannian case, Gromov asserts that
$k_M<\frac{3}{2}n-1$ \cite{gromov}.  More generally, Gilkey and
Nik\v{c}evi\'{c}\cite{GiNi04} have shown that $k_{a,b} \geq
\min(a,b)$.  It is known that there are no proper $\CH_1$ Riemannian
manifolds in 4 dimensions \cite{SeSuVa92}.  In the 3-dimensional,
Lorentzian case, proper $\CH$ geometries have been classified
\cite{Bueken97,KoVl98} and it is known that $k_{1,2}=2$ \cite{BuDj00}.
In the 4-dimensional, Lorentzian case proper $\CH_1$ manifolds were
shown to exist in \cite{BuVa97}.  The recent book by Gilkey
\cite{gilkey} has additional references, and examples of
higher-dimensional proper $\CH$ manifolds of general signature.

Regarding the quantity $k_{1,3}$, Gilkey has conjectured that
$k_{1,3}=2$, and more generally that $k_{a,b} = \min(a,b)+1$
\cite{gilkey}.  However, in the present paper, we establish that these
conjectures are false by showing that $k_{1,3}=3$.  We do this by
proving that in the 4-dimensional, Lorentzian case a proper $\CH_3$
metric does not exist, and by classifying and exhibiting all proper
$\CH_2$ metrics. Equations \eqref{eq:ch2om1}-\eqref{eq:ch2om4} give
the proper $\CH_2$ exact solution as a null-orthogonal tetrad. All
proper $\CH_2$ spacetimes belong to this family, which depends on two
essential constants and one function of one variable.  We also show
that this family is a specialization of type N exact solutions for
coupled gravity and electromagnetic waves propagating in anti-de
Sitter background, first described in \cite{GP81} and \cite{orr}.
Further analysis reveals that, generically, there are no Killing
vectors, but that there is a singular subcase with an $\SL_2\Rset$
isometry group and another singular subfamily with a 1-dimensional
isometry group.  Finally, we prove that the generic, proper $\CH_2$
metrics satisfy the Collins--D'Inverno conditions and enjoy the
remarkable property of $q_M=7$.

\subsection{The CH field equations}
A methodology for expressing and analyzing the field equations for a
$\CH$ geometry is essential to our investigation.  Previously,
Estabrook and Wahlquist described vacuum solutions\cite{EsWa89} as an
involutive exterior differential system( EDS) on the bundle of
second-order frames.  Our approach is to formulate the necessary field
equations as an EDS using two-forms and commutator quantities
(equivalently, connection components) as canonical variables.  Unlike
the field equations for vacuum, the field equations for $\CH$
spacetimes are, in general, overdetermined, with integrability
condition that manifest as algebraic constraints on the curvature and
connection scalars.  Our result is proved by deriving integrable
configurations for the $\CH$ field equations corresponding to various
algebraic types of the curvature tensor, and by using Singer's
criterion to rule out the homogeneous subcases.  We will use this
method to classify proper $\CH_1$ four-dimensional, Lorentz geometries
in a forthcoming publication.

Section 2 of the present paper introduces the necessary field
variables required to formulate the CH field equations.  Section 3
recasts these $\CH$ equations as an EDS, and introduces the crucial
concept of a $\CH$-configuration, the algebraic data that underlies a
$\CH$ geometry.  The actual classification and the proof of our main
result is found in Section 4.  The final section contains some
concluding remarks.

\section{The $\CH$ equations}
\subsection{Preliminaries}
A homogeneous space is fully described by the structure constants of
the underlying Lie algebra.  These constants satisfy algebraic
constraints coming from the Jacobi identity.  Similarly, every
curvature-homogeneous manifold is associated with a collection of
constants and field variables that satisfy algebraic and differential
constraints imposed by the NP equations (2nd structure equations) and
Bianchi identities.

Let $x^i$ be a system of local coordinates on an $n$-dimensional
manifold $M$.  Let $\eta_{ab}=\eta_{ba}$ be a constant inner product of a fixed
signature on an $n$-dimensional vector space $V\cong \Rset^n$.  We are
interested in the case of $n=4$ and of Lorentzian signature, but much
of the underlying theory can be given without these assumptions.
Henceforth, $i,j=1,\ldots, n$ are coordinate indices and
$a,b,c=1,\ldots, n$ are frame indices.  We use $\eta_{ab}$ to lower
and raise frame indices as needed.  Complex conjugation will be
denoted by an asterisk superscript.

Let $\be_a$ be a tetrad/frame, and $\bom^a$ be the dual coframe on $M$.
Let $y^a{}_i$ denote the covariant frame components.  Thus,
\begin{eqnarray}
  \label{eq:yaidef}
  &&\partial_i = \frac{\partial}{\partial x^i} = y^a{}_i \be_a,\\
  \label{eq:bomdef}
  &&\bom^a = y^a{}_i dx^i,\\
  \label{eq:metricdef}
  &&g_{ij} = y^a{}_i y^b{}_j \,\eta_{ab}.
\end{eqnarray}
Let $K^a{}_{bc}=-K^a{}_{cb}$ denote the commutator quantities (structure
functions):
\begin{eqnarray}
  &&[\be_b,\be_c] = K^a{}_{bc}\, \be_a,\\
  \label{eq:dualcommutator}
  &&\rmd\bom^a = - \frac{1}{2} K^a{}_{bc}\, \bom^b\wedge \bom^c,\\
  \label{eq:yaifieldeq}
  && y^a{}_{[i,j]} = \frac{1}{2} K^a{}_{bc}\, y^b{}_i\, y^c{}_j.
\end{eqnarray}
Let
\begin{eqnarray}
  \label{eq:Gdef}
  && G=\O(\eta)= \{ X^a{}_b : X_{ca} X^c{}_b = \eta_{ab}\},\\
  && \fg=\o(\eta)= \{ A^a{}_b : A_{(ab)} = 0 \}
\end{eqnarray}
denote, respectively, the 
\[N:=n(n-1)/2\]
 dimensional group of
$\eta$-orthogonal transformations and the corresponding Lie algebra of
skew-symmetric infinitesimal transformations.  Let
\[\bA_{\alpha}=(A^a{}_{b\alpha}),\quad
\bTheta^\alpha= (\Theta_a{}^{b\alpha}),\]
be a basis of $\fg$ and the dual
basis, respectively, and let $\C^\alpha_{\beta\gamma}$ be the corresponding
$\fg$-structure constants.  Thus,
\begin{eqnarray}
  &&[\bA_\alpha,\bA_\beta] = C^\gamma{}_{\alpha\beta}\, \bA_\gamma,\\
  && A^a{}_{c\alpha} A^c{}_{b\beta} - A^a{}_{c\beta} A^c{}_{b\alpha} =
  C^\gamma{}_{\alpha\beta} A^a{}_{b\gamma},\\
  &&A_{(ab)\alpha} = 0,\quad \Theta^{(ab)\alpha} = 0.
\end{eqnarray}
Henceforth, $\alpha,\beta,\gamma=1,\ldots, N$ denote $\fg$-indices (in
effect, these are bivector indices.)  Let $\Gamma^\alpha{}_a$
denote the connection scalars projected onto this basis.  These are
linearly equivalent to the commutator quantities:
\begin{eqnarray}
  \label{eq:KabcGammadef}
  &&K^a{}_{bc} = \Gamma^\alpha{}_{[c}A^a{}_{b]\alpha},\\
  &&\Gamma^\alpha{}_c = \Theta^{ab\alpha}(2K_{[ab]c} - K_{cab}).
\end{eqnarray}
Respectively, let
\begin{eqnarray}
  \label{eq:bGdef}
  &&\bG^\alpha= \Gamma^\alpha{}_c\, \bom^c, \\
  \label{eq:bOmdef}
  &&\bOm^\alpha = \rmd \bG^\alpha +  \frac{1}{2} C^\alpha{}_{\beta\gamma}
  \bG^\beta \wedge \bG^\gamma = \frac{1}{2} R^\alpha{}_{bc}\, \bom^b\wedge \bom^c,
\end{eqnarray}
be the $\fg$-valued connection 1-form and the curvature 2-form.  The
commutator equation \eqref{eq:dualcommutator} can now be rewritten as
\begin{equation}
  \label{eq:dualcommutator1}
  d\bom^a = -A^a{}_{b\alpha} \bG^\alpha\wedge \bom^b.
\end{equation}
Above,
\begin{eqnarray}
  \label{eq:NPcomp}
  R^\alpha{}_{ab}&&=
  2\Gamma^\alpha{}_{[b,a]}+C^\alpha{}_{\beta\gamma}
  \Gamma^\beta{}_{[a}\Gamma^\gamma{}_{b]}-2 \Gamma^\alpha{}_c
  \Gamma^\beta{}_{[a} A^c{}_{b]\beta} \\
  &&=2\Gamma^\alpha{}_{[b,a]}-C^\alpha{}_{\beta\gamma}
  \Gamma^\beta{}_{[a}\Gamma^\gamma{}_{b]}-2 (\bA_\beta\cdot
  \Gamma)^\alpha{}_{[a} \Gamma^\beta{}_{b]},
\end{eqnarray}
denote the curvature scalars,
and
\begin{equation}
  \label{eq:gammaaction}
  (\bA_\beta\cdot \Gamma)^\alpha{}_a = C^\alpha{}_{\beta\gamma}
  \Gamma^\gamma{}_a - \Gamma^\alpha{}_c A^c{}_{a\beta}
\end{equation}
denotes the action of $\fg$ on $\Lin(V,\fg)$.  The curvature
scalars obey the algebraic and differential Bianchi
identities;  respectively,
\begin{eqnarray}
  \label{eq:abianchicomp}
  && R^\alpha{}_{[bc}A^a{}_{d]\alpha}  = 0,\\
  \label{eq:dbianchicomp}
  && R^\alpha{}_{[ab,c]} = -(\bA_\beta\cdot R)^\alpha{}_{[ab}
  \Gamma^\beta{}_{c]}, 
\end{eqnarray}
where
\begin{equation}
  \label{eq:Raction}
  (\bA_\beta \cdot R)^\alpha{}_{ab} = C^\alpha{}_{\beta\gamma}
  R^\gamma{}_{ab} + 2 R^\alpha{}_{c[a} A^c{}_{b]\beta}
\end{equation}
denotes the $\fg$-action on $\Lin(\Lambda^2 V,\fg)$.

\subsection{The $\CH$ data}
By definition, a $\CH_k$ manifold is specified by an array of
constants
\[ \tR^{(k)} = (\tR^0, \tR^1,\ldots, \tR^k) =  (\tR^\alpha{}_{ab},
\tR^\alpha{}_{abc}, \ldots 
\tR^\alpha{}_{abc_1\ldots c_k})\]  
such that
\begin{equation}
  \label{eq:chi}
  \nabla_{c_1\ldots c_i} R^\alpha{}_{ab} = \tR^\alpha{}_{abc_1\ldots
    c_i},\quad i=0,1,\ldots,k,
\end{equation}
relative to some $\eta$-orthogonal frame.  Note: henceforth a tilde
decoration denotes an array of constants.  

There are two important observations to be made at this point.  First,
it is more efficient to represent the algebraic $\CH$ data in terms of
connection scalars rather than curvature scalars.  To that end, set
\begin{equation}
  G_{-1} := G,\qquad  \fg_{-1} :=\fg,
\end{equation}
and let $G_i\subset G_{i-1}, \; i=0,1,\ldots, k$ denote the subgroup
that leaves invariant
\[\tR^{(i)} := (\tR^0, \tR^1,\ldots, \tR^i).\]
Let $\fg_i$ denote the corresponding Lie algebra, and set
\begin{equation}
  N_i:=\dim\fg_i,\quad  \hN_i:= N-N_{i},\quad i=0,\ldots, k+1.
\end{equation}
Arrange the basis of $\fg$ into $k+2$ groups of generators,
\begin{equation}
  \bA_{\rho_1},\ldots, \bA_{\rho_k},\bA_\lambda,  \bA_\xi,
\end{equation}
where 
\begin{equation}
  \label{eq:indexscheme3}
  \bA_\xi,\quad \hN_{k}+1\leq \xi \leq N
\end{equation}
is a basis of $\fg_k$, where
\begin{equation}
  \label{eq:indexscheme2}
  \bA_\lambda, \bA_\xi,\quad \hN_{k-1}+1\leq \lambda \leq \hN_{k}
\end{equation}
is a basis of $\fg_{k-1}$ and where
\begin{equation}
  \label{eq:indexscheme1}
 \bA_{\rho_i},\ldots, \bA_{\rho_k}, \bA_\lambda, \bA_\xi,\quad
  \hN_{i-2}+1\leq \rho_i
  \leq \hN_{i-1}
\end{equation}
is a basis of $\fg_{i-1},\; i=1,\ldots, k$. 
Henceforth, we restrict the indices $\xi,\lambda, \rho_i$ to the
ranges indicated above, and use the Einstein convention to sum over
these indices. 

By the usual formula for the covariant derivative,
\begin{equation}
  \nabla_c R^\alpha{}_{ab}\, \bom^c =\rmd R{}^\alpha{}_{ab} +
  (\bA_\beta \cdot   R)^\alpha{}_{ab}\,\bG^\beta,
\end{equation}
where the second term on the right is defined in \eqref{eq:Raction}.
In a $\CH_1$ geometry, $R^\alpha{}_{ab} = \tR^\alpha{}_{ab}$ is
constant, and since $\fg_0$ leaves invariant the latter array, we
actually have
\begin{equation}
  \label{eq:DR1}
  \tR^\alpha{}_{abc} = (\bA_{\rho_1} \cdot
  \tR)^\alpha{}_{ab}\,\Gamma^{\rho_1}{}_c.
\end{equation}
The scalars $\Gamma^{\rho_1}{}_a$ specify an element of
$\Lin(V,\fg/\fg_0)$.  By definition of $\fg_0$, the linear map
$(\Gamma^{\rho_1}{}_a)\mapsto(\tR^\alpha{}_{abc})$ has a trivial
kernel. Hence, it is possible to solve the linear system
\eqref{eq:DR1} and express $\Gamma^{\rho_1}{}_a$ in terms of
$\tR^\alpha{}_{ab}$ and $\tR^\alpha{}_{abc}$ --- rational in the
former, and linear in the latter. Therefore, in a $\CH_1$ context
$\Gamma^{\rho_1}{}_a = \tG^{\rho_1}{}_a$ is an array of constants.
The following Proposition makes this more precise.
\begin{proposition}
  \label{prop:gammaaction}
  Let $\fg$ be a finite-dimensional Lie algebra, $V$ a $\fg$-module,
  and $T$ the tensor algebra over $V$.  Let us define a
  bilinear product on $\fg\otimes T$ by setting
  \begin{equation}
    \label{eq:curvproduct}
    (a\otimes \alpha) \cdot (b\otimes \beta) := [a,b] \otimes
    \beta\otimes\alpha + b \otimes (a\cdot \beta) \otimes
    \alpha,\quad a,b\in \fg,\quad \alpha,\beta\in T.
  \end{equation}
  This product satisfies the Leibniz rule with respect to the action
  of $\fg$ on $\fg\otimes  T$.
  In other words,  for $a,b,c\in \fg$ and $\beta,\gamma\in
  T$, we have
  \begin{equation}
    \label{eq:gtuleib}
    a\cdot((b\otimes\beta) \cdot (c\otimes \gamma)) = ([a,b]\otimes
    \beta + b\otimes(a\cdot \beta)) \cdot (c\otimes \gamma)+ (b\otimes
    \beta)\cdot([a,c]\otimes \gamma + c\otimes (a\cdot \gamma)).
  \end{equation}
\end{proposition}
\begin{proof}
  The left-hand side of \eqref{eq:gtuleib} expands to
  \begin{eqnarray*}
    \mbox{LHS} &=& [a,[b,c]] \otimes \gamma\otimes
  \beta + [b,c]\otimes 
  (a\cdot \gamma)\otimes \beta +[b,c]\otimes \gamma\otimes (a\cdot \beta) +\\
  &&  +[a,c]\otimes (b\cdot \gamma)\otimes \beta + c\otimes (a\cdot
  b\cdot \gamma)\otimes \beta + c\otimes (b\cdot \gamma)\otimes
  (a\cdot \beta)\\
  &=& [[a,b],c]\otimes \gamma\otimes \beta + c \otimes ([a,b]\cdot
  \gamma) \otimes \beta + [b,c]\otimes \gamma\otimes (a\cdot \beta) +
  c\otimes (b\cdot \gamma)\otimes (a\cdot \beta)+\\
  && +[b,[a,c]] \otimes \gamma\otimes \beta + [b,c] \otimes
  (a\cdot \gamma)\otimes \beta + [a,c]\otimes (b\cdot \gamma) \otimes
  \beta + c\otimes (b\cdot a\cdot \gamma)\otimes \beta.
  \end{eqnarray*}
  By inspection, the latter is equal to the right-hand side of
  \eqref{eq:gtuleib}.
\end{proof}
Henceforth, let us set
\[ \tG^{(1)} := (\tG^{\rho_1}{}_a) \] and use \eqref{eq:curvproduct}
to rewrite \eqref{eq:DR1} as
\[ \tR^1 = \tG^{(1)} \cdot \tR^0.\] 
Let $A\in \fg_0$; i.e.  $A\cdot
\tR^0=0$.  By Proposition
\ref{prop:gammaaction},
\begin{equation}
  A \cdot \tR^1 = (A\cdot \tG^{(1)}) \cdot \tR^0 + \tG^{(1)} \cdot
  (A\cdot \tR^0) = (A\cdot \tG^{(1)}) \cdot \tR^0.
\end{equation}
The above identity also establishes that $A\in\fg_1$ if and only if
$A\cdot \tG^{(1)} = 0$.  Hence, $\fg_1$ can be characterized as the
automorphism subalgebra of $\tG^{(1)}$.  Therefore, a $\CH_1$ geometry
is fully described by the constants $\tR^0,\tG^{(1)}$.


In a $\CH_k$ context, formula \eqref{eq:DR1} extends to covariant
derivatives of higher  
order:
\begin{equation}
  \label{eq:DRk}
   \tR^\alpha{}_{abc_1\cdots c_i}  = \sum_{j=1}^i 
   (\bA_{\rho_j}\cdot  
  \tR)^\alpha{}_{abc_1\cdots c_{i-1}} \tG^{\rho_j}{}_{c_i},\quad
  i=1,\ldots, k.
\end{equation}
Setting
\begin{equation}
  \label{eq:tGammadef}
  \tGamma^{(i)} = \left(\begin{array}{c}
      \tG^{\rho_1}{}_a\\ \vdots \\ \tG^{\rho_i}{}_a
      \end{array}\right)
\end{equation}
the equation \eqref{eq:DRk} can be expressed, symbolically, as
\begin{equation}
  \tR^{i} = \tG^{(i)} \cdot \tR^{i-1}, \quad i =1,\ldots, k.
\end{equation}
Therefore, a $\CH_k$ geometry is fully described by the constants
$\tR^{0}, \tG^{(k)}$.  

\subsection{Proper $\CH$ geometry}
The second crucial observation is that the condition that
distinguishes proper $\CH$ geometries from homogeneous geometries can
be restated in terms of the automorphism subalgebras $\fg_i$,
c.f. Theorem \ref{thm:singcrit}.
\begin{theorem}[Singer's criterion, restated]
  \label{thm:SC}
  If $M$ is a \textbf{proper} $\CH_k$ manifold, then, necessarily,
  $\fg_k \subsetneq \cdots \subsetneq \fg_0 \subsetneq \fg_{-1}$ is a
  chain of proper inclusions.
\end{theorem}
This follows from Karlhede's algorithm.  In a $\CH$ manifold, if
$\fg_i = \fg_{i-1}$, then the algorithm terminates because $t_i =
t_{i-1}=0$.  Since all differential invariants are constants, the
manifold is a homogeneous space. The constants $\tR^{(k)}$ are
invariants that define the structure constants of the corresponding
Lie algebra of Killing vectors. For more details, see \cite{singer}
and Chapter 2.6 of \cite{PoSp}.

Hence, if the geometry is properly $\CH_k$, i.e., if it is not
$\CH_{k+1}$, then, at the $(k+1)$st order, we have
\begin{equation}
  \fl \nabla_{c_{k+1}\cdots c_1} R^\alpha{}_{ab} = 
  (\bA_\lambda\cdot\tR)^\alpha{}_{abc_1\cdots c_k}
  \Ga^\lambda{}_{c_{k+1}}
  +\sum_{i=1}^k
  (\bA_{\rho_i}\cdot\tR)^\alpha{}_{abc_1\cdots c_k} \tG^{\rho_i}{}_{c_{k+1}},
\end{equation}
where not all $\Gamma^\lambda{}_a$ are constants.  Symbolically, we
will express this as
\[ R^{k+1} = \Gamma^{(k+1)} \cdot \tR^k,\]
where
\[ \Gamma^{(k+1)} :=\left(\begin{array}{c} \tG^{(k)} \\
    \Gamma^{\lambda}{}_a
  \end{array}\right).
\]


\subsection{Transformations of the $\CH$ data}
A $\CH_k$ metric does not determine the groups $G_i$ and the constants
$\tR^0, \tG^{(k)}$ uniquely, but only up to a certain transformation.
The general transformation law $\Gamma^\alpha{}_a \mapsto
\hat{\Gamma}^\alpha{}_a$ for connection scalars involves derivatives:
\begin{equation}
  \label{eq:cocompxform}
  \hat{\Gamma}{}^\alpha{}_a \bom^a = (X \cdot \Gamma)^\alpha{}_a \bom^a +
   (X^{-1} dX)^\alpha.
\end{equation}
Here, $X$ is a $G$-valued function on $M$, and  $X\cdot \Gamma$
denotes the G-action on $\Lin(V,\fg)$.  Note that
$(\Gamma^\lambda{}_a)$ is a field taking values in
$\Lin(V,\fg_{k-1}/\fg_k)$.  Hence, if we restrict the values of
the frame transformation to $G_k$, i.e., $X:M\to G_k$, then $X^{-1}
dX$ takes values 
in $\fg_k$, and the transformation law for the connection components
modulo $\fg_k$ becomes tensorial:
\begin{eqnarray}
  \label{eq:Grxlaw}
  \hat{\Gamma}^\lambda{}_a = (X\cdot \Gamma)^\lambda{}_a.
\end{eqnarray}
This makes sense, because the scalars $\Gamma^\lambda{}_a$ depend
linearly on $R^{k+1}$, and the
components of the latter transform tensorially.

Thus, in a $\CH_k$ manifold the group $G_0$ is only determined up to
an $G_{-1}$ conjugation.  If $X\in G_{-1}$ is a constant frame
transformation, i.e., if $dX=0$, then the corresponding frame
transformation leaves $\tR^0, \tG^{(k)}$ constant.  If $G_0$ is fixed,
then the constants $\tR^0$ are determined up to a choice of
conjugation by a constant $X\in\N(G_0$), where the latter denotes the
normalizer of $G_0$.  More generally, once $G_{i},\; i=1,\ldots, k-1$
is fixed, then $G_{i+1}$ and the constants $\tG^{(i+1)}$ are
determined up to conjugation by a constant frame transformation
$X\in\N(G_0)\cap\cdots\cap\N(G_{i})$.  The latter is the group that
preserves the chain $G_i\subset \cdots \subset G_0\subset G_{-1}$.  Once $G_k$
has been fixed, the constant data is fixed.  However, the scalars
$\Gamma^\lambda{}_a$ obey a $G_k$ transformation law
\eqref{eq:Grxlaw}, and can be normalized using a
non-constant frame transformation $X:M\to G_k$.

\subsection{The $\CH$ constraints}
The $\CH_k$ condition imposes certain algebraic and differential
constraints on the constants $\tR^0=(\tR^\alpha{}_{ab}),
\tG^{(k)}=(\tG^{\rho_i}{}_a)_{i=1}^k $ and field variables
$\Gamma^\lambda{}_a$.  To express these, we introduce the following
quantities:
\begin{eqnarray}
  \label{eq:Xiadef}
  \txi^a{}_{bcd} &:=&
  \tR^\alpha{}_{\vphantom{\alpha}[bc}A^a{}_{d]\alpha}\\ 
  \label{eq:Xialdef}
  \fl\quad \tXi^{\alpha}{}_{abc} &:=& (\bA_{\rho_1}\cdot \tR)^\alpha{}_{[ab}
    \tG^{\rho_1}{}_{c]}\\
  \label{eq:Updef1}
  \fl \quad \tUp^{\rho_i}{}_{ab} &:=& \tR^{\rho_i}{}_{ab}-
  \sum_{\sigma,\tau=1}^{\hN_{i-1}}   
  C^{\rho_i}{}_{\sigma\tau} 
  \tG^{\sigma}{}_a \tG^{\tau}{}_b +2\sum_{\sigma=1}^{\hN_{i-1}}
  \tG^{\rho_i}{}_c 
  \tG^{\sigma}{}_{[a} A^c{}_{b]\sigma}
  +\\ \nonumber
  &&+2(\bA_{\rho_{i+1}}\cdot
  \tG)^{\rho_i}{}_{[a}\tG^{\rho_{i+1}}{}_{b]} ,\quad i=1,\ldots,k-1;\\
  \label{eq:Updef2}
  \Upsilon^{\rho_k}{}_{ab} &:=& \tR^{\rho_k}{}_{ab}-
  \sum_{\sigma,\tau=1}^{\hN_{k-1}}   
  C^{\rho_i}{}_{\sigma\tau} 
  \tG^{\sigma}{}_a \tG^{\tau}{}_b +2\sum_{\sigma=1}^{\hN_{k-1}}
  \tG^{\rho_k}{}_c 
  \tG^{\sigma}{}_{[a} A^c{}_{b]\sigma}
  +\\ \nonumber
  &&+2(\bA_\lambda \cdot \tG)^{\rho_k}{}_{[a}
  \Gamma^\lambda{}_{b]} ;\\
  \label{eq:Updef3}
  \fl \quad\Upsilon^\lambda{}_{ab} &:=&
    \tR^\lambda{}_{ab}-\sum_{\sigma,\tau=1}^{\hN_{k-1}} 
  C^\lambda{}_{\sigma\tau}\tG^\sigma{}_a \tG^\tau{}_b  
  + 2 \sum_{\sigma=1}^{\hN_{k-1}} ( \Gamma^{\lambda}{}_{c\hphantom{[} }
  \tG^\sigma{}_{[a} A^c{}_{b]\sigma}-
  \Gamma^\mu{}_{a}\tG^\sigma{}_{b}C^\lambda{}_{\mu\sigma}  ) \\ \nonumber
  &&-  C^\lambda{}_{\mu\nu}\Gamma^\mu{}_a \Gamma^\nu{}_b  + 2\,
  \Gamma^{\lambda}{}_{c\hphantom{[} }
  \Gamma^\mu{}_{[a} A^c{}_{b]\mu} , 
\end{eqnarray}
where in \eqref{eq:Updef3} $\mu,\nu$ have the same range as $\lambda$,
as per \eqref{eq:indexscheme2}.  The algebraic and differential
Bianchi identities  
\eqref{eq:abianchicomp} \eqref{eq:dbianchicomp} reduce to the
following  constraints:
\begin{eqnarray}
  \label{eq:algbianchi}
  &&\txi^a{}_{bcd} = 0,\\
  \label{eq:diffbianchi}
  &&\tXi^\alpha{}_{bcd} = 0.
\end{eqnarray}
The NP equations \eqref{eq:NPcomp} corresponding to generators
$\bA_{\rho_1},\ldots, \bA_{\rho_k}$ also reduce to the following
algebraic constraints
\begin{eqnarray}
  \label{eq:NPtieri}
  \tUp^{\rho_i}{}_{ab} = 0,\quad  i=1,\ldots, k-1,\\   
  \label{eq:NPtierk}
  \Upsilon^{\rho_k}{}_{ab}=0.
\end{eqnarray}
The NP equations corresponding to generators $\bA_\lambda$ give
$\CH_k$ field equations:
\begin{eqnarray}
  &&d\Gamma^\lambda{}_b \wedge \bom^b+
  (\bA_\xi\cdot \Gamma)^\lambda{}_b\,  \bG^\xi \wedge \bom^b =
  \frac{1}{2}\Upsilon^\lambda{}_{ab} \, \bom^a\wedge \bom^b ,\\
  \label{eq:lambdafieldeq}
  &&\Gamma^\lambda{}_{[a,b]} =   (\bA_\xi\cdot \Gamma)^\lambda{}_b
  \Gamma^\xi{}_a -\frac{1}{2}\Upsilon^\lambda{}_{ab}.
\end{eqnarray}
The field equations for the scalars $\Gamma^\xi{}_a$ are similar.
Note that if $G_k$ is trivial, as happens in the $\CH_2$ examples
derived in Section \ref{sect:ch2}, then \eqref{eq:lambdafieldeq}
becomes, simply
\begin{equation}
  \label{eq:lambdafieldeq2}
  \Gamma^\lambda{}_{[a,b]} = -\frac{1}{2} \Upsilon^\lambda{}_{ab}.
\end{equation}

\subsection{Integrability conditions}
Equations \eqref{eq:NPtieri} are polynomial constraints on the
constants $\tR^{0}, \tG^{(k)}$.  Equations \eqref{eq:NPtierk} are
linear algebraic constraints while \eqref{eq:lambdafieldeq} are
quasi-linear differential equations in the field variables
$\Gamma^\lambda{}_a$.  The scalars $\Gamma^\xi{}_a$ are not subject to
any algebraic constraints.  The equations \eqref{eq:NPtieri}
\eqref{eq:NPtierk} \eqref{eq:lambdafieldeq} are necessary conditions,
but not, in general, sufficient to describe a $\CH_k$ geometry because
of the presence of integrability conditions.  The complication,
roughly speaking, is that the derivatives of the algebraic
constraints \eqref{eq:NPtierk} together with differential constraints
\eqref{eq:lambdafieldeq} may imply additional zero-order (algebraic)
constraints (obtained by eliminating all first-order terms.)  The
derivatives of these zero-order constraints may imply further
algebraic constraints, etc.  In addition to zero-order integrability
conditions, equations \eqref{eq:lambdafieldeq} may fail to be
involutive because of first order obstructions.  Taking derivatives of
\eqref{eq:lambdafieldeq} yields second-order differential equations in
$\Gamma^\lambda{}_a$.  It is conceivable that a particular linear
combination of these prolonged second-order equations eliminates all
second-order derivatives, and furnishes additional first-order
constraints that are independent of \eqref{eq:lambdafieldeq}.


Fix $\eta_{ab}$ of the desired signature.  Henceforth, $x^i, y^a{}_i,
\Gamma^\alpha{}_a$ are canonical coordinates.  Let $\F M$ denote the
$\mathrm{GL}_n\Rset$ frame bundle over $M$.  The variables $x^i, y^a{}_i$ are
canonical coordinates on $\F M$, while the variables
$\Gamma^\alpha{}_a$ are canonical coordinates on the vector space
$\Lin(V,\fg)$.  The $x^i$ are independent variables, while
\begin{eqnarray}
  &y^a{}_i = y^a{}_i(x^1,\ldots,x^n),\\
  &\Gamma^\alpha{}_i = \Gamma^\alpha{}_i(x^1,\ldots,x^n)
\end{eqnarray}
are the dependent variables.  The metric $g_{ij}$, as given by
\eqref{eq:metricdef}, is a derived dependent variable.  The
differential forms $\bom^a,\bG^\alpha, \bOm^\alpha$ defined,
respectively, by \eqref{eq:bomdef}\eqref{eq:bGdef} \eqref{eq:bOmdef}
should also be regarded as unknown quantities.

Let
\begin{eqnarray}
  &&\tR^{0}=(\tR^\alpha{}_{ab}) \in \Lin(\Lambda^2 V,    \fg),\\
  \label{eq:tGj}
  &&\tG^{(j)}=(\tG^{\rho_i}{}_a)_{i=1}^j \in
  \Lin(V,\fg_{j-2}/\fg_{j-1}),\quad j=1,\ldots, k
\end{eqnarray}
be arrays of constants that satisfy \eqref{eq:algbianchi}
\eqref{eq:diffbianchi} \eqref{eq:NPtieri}. The first set of
constraints comes from algebraic Bianchi relations, the second set
from differential Bianchi, and the third set from tier 1 through $k-1$
NP equations.  Let us again emphasize that in a $\CH_k$ context some
differential constraints reduce to purely algebraic constant constraints. In
\eqref{eq:tGj}, the automorphism Lie algebras are defined inductively
by
\begin{eqnarray}
  && \fg_{-1} = \fg,\\
  && \fg_0  = \Aut\, \tR^0=\{ A\in \fg_{-1} : (A\cdot \tR)^\alpha{}_{ab} = 0\},\\
  && \fg_i = \Aut\, \tR^{(i)} \\ \nonumber
  && \quad = \Aut \, \tR^0 \cap \Aut \, \tR^1 \cap
  \cdots \cap \Aut \, \tR^i \\ \nonumber
  && \quad = \Aut\, \tR^0\cap \Aut\, \Gamma^{(i)}\\ \nonumber
  && \quad = \Aut\{ A \in
  \fg_{i-1} : (A\cdot \tG)^{\rho_i}{}_a = 
  0\},\quad i=1,\ldots, k
\end{eqnarray}
Let $Z\subset
\Lin(V,\fg)$ be an algebraic variety defined by the equations
\begin{eqnarray}
  &&\Gamma^{\rho_i}{}_a = \tG^{\rho_i}{}_a,\quad i=1,\ldots, k,
\end{eqnarray}
by the linear equations \eqref{eq:NPtierk}, and by some additional
polynomial equations in the $\Gamma^\lambda{}_a$.   
\begin{definition}
  \label{def:chconfig}
  We will call the pair $(\tR^0,Z)$ a $\CH_k$ configuration.  We will
  say that the configuration is \emph{proper}, if every inclusion
  $\fg_i\subsetneq \fg_{i-1}$ is proper. We will say that two
  configurations are \emph{equivalent} if they can be related by a constant
  $\O(\eta)$ conjugation.
\end{definition}
Let $ZM = \F M \times Z$.
Set
\begin{eqnarray}
  &&\btau^a := \rmd\bom^a - A^a{}_{b\alpha}\bG^\alpha,\\
    &&\bUp^\alpha := \bOm^\alpha - \frac{1}{2} \tR^\alpha{}_{ab}\,
  \bom^a\wedge \bom^b,
\end{eqnarray}
and let $(\cI,\bTh)$ be the exterior differential system\cite{IvLa03}
\cite[Chapter 18]{olver} on $Z M$
generated by the 2-forms $\btau^a, \bOm^\lambda, \bOm^\xi$, subject to
the independence condition
\begin{equation}
  \bTh=\rmd x^1\wedge \cdots \wedge\rmd x^n\neq 0.
\end{equation}
Equivalently, as per \eqref{eq:yaifieldeq} \eqref{eq:KabcGammadef}, we
may consider scalar equations 
\begin{equation}
  y^a{}_{[i,j]} = \frac{1}{2} \Gamma^\alpha{}_{[c}A^a{}_{b]\alpha},
\end{equation}
and NP equations \eqref{eq:lambdafieldeq} constrained by the variety $Z$.
\begin{definition}
  \label{def:chconfig1}
  We will say that a $\CH_k$ configuration is \emph{free of torsion}
  if all zero order integrability constraints are satisfied
  identically on $ZM$, i.e., if there exists an $n$-dimensional
  integral element of $(\cI,\bTh)$ above every point of $Z M$.
\end{definition}

\begin{proposition}
  The exterior ideal $\cI$ is differentially closed.
\end{proposition}
\begin{proof}
  By \eqref{eq:bOmdef},
  \[ \bUpsilon^\alpha = \rmd \bG^\alpha +  \frac{1}{2} C^\alpha{}_{\beta\gamma}
  \bG^\beta \wedge \bG^\gamma - \frac{1}{2} \tR^\alpha{}_{ab}\,
    \bom^a\wedge \bom^b
  \]
  Since $\Gamma^{(k)}$ are constants, we have
  \[ \bUpsilon^{\rho_i} = \frac{1}{2} \Upsilon^{\rho_i}{}_{ab}
  \bom^a\wedge \bom^b = 0,\quad i=1,\ldots, k\] by definition.
  Therefore, $\bUpsilon^\alpha\in \cI$ for all $\alpha$.  As well, the
  following identities hold:
  \begin{eqnarray} 
    &&\rmd  \btau^a \equiv \frac{1}{6}\, \xi^a{}_{bcd}\, \bom^a \wedge \bom^b
    \wedge \bom^c \pmod {\cI},\\
    &&\rmd \bUpsilon^\alpha \equiv \frac{1}{6}\, \Xi^\alpha{}_{abc} \,
    \bom^a \wedge \bom^b \wedge \bom^c \pmod{\cI},
  \end{eqnarray}
  with $\xi^a{}_{bcd}, \Xi^\alpha{}_{bcd}$ defined in
  \eqref{eq:Xiadef} \eqref{eq:Xialdef}.  Again, by definition, these
  polynomials vanish on $Z$.  Therefore, the three-forms $\rmd
  \btau^a, \rmd\bUpsilon^\alpha$ all belong to
  $\cI_\cR$.
\end{proof}

In and of itself, the above result does not guarantee involutivity
because of the potential presence of zero and first order
integrability constraints.  However, in light of the above result, the
construction of proper $\CH_k$ geometries is reduced to the search for
proper, torsion-free configurations up to $G$-equivalence.  After
classifying proper, torsion-free $\CH_k$ configurations, all that
remains is to test these configurations for involutivity, that is the
absence of additional 1st-order integrability conditions. Below, we
apply this approach to classify all proper $\CH_2$ Lorentzian
four-manifolds, and to prove the non-existence of proper $\CH_3$
Lorentzian four-manifolds.

The algebraic constraints \eqref{eq:NPtierk} and the differential
constraints \eqref{eq:lambdafieldeq} are both consequences of the NP
equations.  An analysis of the branching arising from the zero-order
integrability conditions implied by these constraints leads to a
classification of proper, torsion-free configurations.  We implement
this program for the case of four-dimensional Lorentzian $\CH_2$
geometries in the next section.

\section{Proper $\CH_2$ geometries}
\label{sect:ch2}
\subsection{Derivation of proper configurations.}
In this section we classify proper $\CH_2$ Lorentzian geometries.  Our
method relies in an essential way on Theorem \ref{thm:SC}, Singer's
criterion.  We begin by classifying proper Lie algebra chains
$\fg_{2}\subsetneq \fg_1 \subsetneq \fg_0 \subsetneq \fg_{-1}$, where
$\fg_{-1}$ is the 6-dimensional Lie algebra of infinitesimal Lorentz
transformations, where $\fg_0$ is the automorphism algebra for some
curvature constants $\tR^0 = (\tR^\alpha{}_{ab})$, and where $\fg_i,\;
i=1,2$ is the automorphism algebra of some tier $i$ connection
constants $\tG^{(i)}$.  By focusing on proper chains, we exclude
homogeneous four-dimensional geometries.  These are classified, in the
Riemannian case, in \cite{patrangenaru}, and in the indefinite
signature case in \cite{felsrenner}.

Henceforth, we assume that $M$ is a four-dimensional, analytic
manifold and that $\eta_{ab}$ is the Lorentzian inner product.  We
will express our calculations using the Newman-Penrose (NP) formalism
\cite[Chapter 2]{ES}, which is based on complexified null tetrads 
\[(
\be_a) = (\vm, \bvm, \vn, \vell) = (\delta, \delta^*,\Delta,  D).\]
The inner product and the metric
are given by
\begin{eqnarray}
  \label{eq:etadef}
  && g_{ij}\, dx^i dx^j = \eta_{ab} \bom^a\bom^b = 2\bom^1\bom^2 -
  2\bom^3\bom^4,
\end{eqnarray}
The connection scalars $\Gamma^\alpha{}_a$ are labeled by the 12
complex-valued NP spin coefficients:
\begin{eqnarray*}
  \hskip 4em-\bom_{14}&= \sigma\bom^1+\rho\bom^2+\tau\bom^3 + \kappa\bom^4;\\
  \hskip 5em\bom_{23} &= \mu\bom^1+\lambda \bom^2+\nu\bom^3 +\pi\bom^4;\\
  -\frac{1}{2}(\bom_{12}+\bom_{34}) &= \beta \bom^1 +
  \alpha\bom^2+\gamma\bom^3+\epsilon\bom^4.
\end{eqnarray*}
The curvature scalars are labeled by the Ricci scalar
$\Lambda=\overline{\Lambda}$, Hermitian Ricci components
$\Phi_{AB}=\overline{\Phi_{BA}},\; A,B=0,1,2$, and complex Weyl
components $\Psi_C ,\; C=0\ldots 4$ according to the usual
Newman-Penrose scheme:
\begin{eqnarray}
  \label{eq:NPcurv1}
  \small
  \bOm_{14}&=&
  \Phi_{01}(\bom^{3}\!\wedge\bom^{4}-\bom^{1}\!\wedge\bom^{2})  +\Psi_1
  (\bom^{1}\!\wedge\bom^{2}+\bom^{3}\!\wedge\bom^{4}) \\ \nonumber
  && -
  \Phi_{02}\bom^{1}\!\wedge\bom^{3}+\Phi_{00}\bom^{2}\!\wedge\bom^{4} 
  +\Psi_0\bom^{1}\!\wedge\bom^{4} -\left(\Psi_2 +
    2\Lambda\right)\bom^{2}\!\wedge\bom^{3} \\
  \label{eq:NPcurv2}
  \small
  \bOm_{23} &=& \Phi_{21}(\bom^{1}\!\wedge\bom^{2}-\bom^{3}\!\wedge\bom^{4})   -
  \Psi_3(\bom^{1}\!\wedge\bom^{2}+ \bom^{3}\!\wedge\bom^{4})\\ \nonumber
  && +\Phi_{22}
  \bom^{1}\!\wedge\bom^{3}-\Phi_{20}
  \bom^{2}\!\wedge\bom^{4}+\Psi_4\bom^{2}\!\wedge\bom^{3} -  
  (\Psi_2 + 2\Lambda)\bom^{1}\!\wedge\bom^{4}\\ 
  \label{eq:NPcurv3}
  \small
   \frac{1}{2}(\bOm_{12} + \bOm_{34}) &=&  \Phi_{11}
  (\bom^{3}\!\wedge\bom^{4}-\bom^{1}\!\wedge\bom^{2})
  +(\Psi_2-\Lambda)(\bom^{1}\!\wedge\bom^{2}+\bom^{3}\!\wedge\bom^{4})
  \\ \nonumber 
  &&- \Phi_{12}\bom^{1}\!\wedge\bom^{3} +
  \Phi_{10}\bom^{2}\!\wedge\bom^{4}+\Psi_1\bom^{1}\!\wedge\bom^{4}-
  \Psi_3\bom^{2}\!\wedge\bom^{3}
\end{eqnarray}

If the Petrov type is I, II, or III, then one can fully fix the frame
by setting $\Psi_0=\Psi_4=0$ and then normalizing $\Psi_1$ or $\Psi_3$
to $1$.  In other words,  $N_1=N_0=0$, and hence, by Singer's
criterion, every $\CH_1$ manifold of Petrov type I, II, or III is a
homogeneous space.  Proper $\CH_1$ Lorentzian manifolds must,
necessarily, be of Petrov type D, N, or O.  

A proper $\CH_2$ configuration, if one exists, requires that $N_0\geq
2$.  Modulo conjugation by a Lorentz transformation, there are only 5
types of curvature tensor for which the automorphism group has
dimension 2 or higher.  The analysis of these 5 cases and their
subcases is detailed below. Only in case 5.2, do we obtain a proper
$\CH_2$ configuration.

A proper $\CH_3$ configuration, if one exists, requires a proper chain
$\fg_3\subsetneq
\fg_2\subsetneq\fg_1\subsetneq\fg_0\subsetneq\fg_{-1}$, and hence
requires that $N_0=\dim\fg_0 \geq 3$.  Thus the search for proper
$\CH_3$ configurations is limited to cases 1,2,3, below. However, for
each of these cases we rule out the existence of a proper $\CH_2$
configuration.  From this it follows that there does not exist a
proper $\CH_3$ configuration, and hence there does not exists a proper
$\CH_3$ Lorentzian four-manifold.

\paragraph{Case 1.} $\fg_0=\so(3)$.  The curvature is that
of a conformally flat perfect fluid.
\begin{eqnarray*}
  \Psi_A= \Phi_{01} =
  \Phi_{02}=\Phi_{12} = 0.\quad \Phi_{00}=\Phi_{22}=2 \Phi_{11}\neq 0.
\end{eqnarray*}
As a basis of $\o(\eta)$ we take
\begin{equation}
  (\bA_\alpha)= (\be^{34}, \be^{14}, \be^{24}, \be^{13}-\be^{14},
  \be^{23} - 
  \be^{24}, \be^{12}),
\end{equation}
where $\be^{ab}=\be^a\wedge \be^b$ is a basic bivector.  Note that, as per the
indexing scheme \eqref{eq:indexscheme1} - \eqref{eq:indexscheme3},
$\bA_4,\bA_5, \bA_6$ are the $\so(3)$ generators.  The dual connection
components are 
\[ (\bG^\alpha) = (\bom_{34}, \bom_{13}+\bom_{14}, \bom_{23}+\bom_{24},
\bom_{13}, \bom_{23}, \bom_{12}).\]
Hence, the tier-1
connection constants are 
\begin{equation*}
  \tG^{(1)} =
  \left(  \begin{array}{cccc} 
    -\beta-\bar{\alpha}& -\alpha-\bar{\beta} &  -2\gamma_1 &
    -2\epsilon_1\\
    -\sigma+\bar{\lambda} & -\rho+\bar{\mu} & -\tau+\bar{\nu} &
    -\kappa+\bar{\pi} \\
    \mu-\bar{\rho} & \lambda-\bar{\sigma} & \nu-\bar{\tau} & \pi-\bar{\kappa}
  \end{array}
\right),
\end{equation*}
where the $1$ and $2$ subscripts indicates the real part and imaginary
part; e.g., $\gamma=\gamma_1+i\gamma_2$. The Bianchi identities
$\tXi^\alpha{}_{abc}=0$ 
imply that all $\tGu^{\rho_1}{}_a=0$. Hence, necessarily, $\fg_1=\fg_0$.
Therefore, this case does not admit a proper $\CH_1$ configuration,
much less a proper $\CH_2$ configuration.

\noindent
\paragraph{Case 2.} $\fg_0=\so(1,2)$. The curvature constants are
\begin{eqnarray*}
  \Psi_A= \Phi_{01} =
  \Phi_{02}=\Phi_{12} = 0.\quad \Phi_{00}=\Phi_{22}=-2 \Phi_{11}\neq 0.
\end{eqnarray*}
This case is quite similar to Case 1. Again, by the Bianchi
identities, $\fg_1=\fg_0$.  This type of curvature tensor does not
admit proper $\CH_1$ configurations.

\noindent
\paragraph{Case 3.}
The curvature is that of an aligned null
radiation field on a conformally flat background:
\begin{eqnarray*}
  \Psi_C=\Phi_{00}=\Phi_{01}=\Phi_{11}=\Phi_{02}=\Phi_{12} = 0,\quad
  \Phi_{22} \neq 0.
\end{eqnarray*}
The automorphism group of the curvature tensor is three-dimensional,
generated by spins and null rotations; the generators are $\bA_4, \,
\bA_5, \, \bA_6$ where
\begin{eqnarray*}
  (\bA_\alpha) = (\be^{34}, \be^{14}, \be^{24}, \be^{12}, \be^{13}+\be^{23},
  \be^{13}-\be^{23}).
\end{eqnarray*}
The tier-1 connection constants are
\[   (\tG^{\rho_1}{}_a) = \left(
  \begin{array}{cccc}
    -\beta-\bar{\alpha}& -\alpha-\bar{\beta} &  -2\gamma_1 &
    -2\epsilon_1\\
    -\sigma & -\rho & -\tau &
    -\kappa \\
    -\bar{\rho} & -\bar{\sigma} &-\bar{\tau} & -\bar{\kappa}
  \end{array}\right)
\]
By  the Bianchi identities, necessarily
\[ \kappa=\sigma=\rho=0,\quad \alpha=\bar{\tau}/2-\bar{\beta}.\] If
$\tau=0$, then $\fg_1=\fg_0$.  By Singer's criterion, this gives a
homogeneous space. So, we assume that $\tau\neq 0$. Conjugating by a
spin (a $G_0$ transformation), as necessary, we assume without loss of
generality that $\tau=\tau_1$ is real, and that $\fg_1$ is
1-dimensional, generated by imaginary null rotations $\bA_6$.  Thus,
\[ (\tGu^{\rho_2}{}_a) = \left(
  \begin{array}{cccc}
    \tau/2-2\beta & -\tau/2 + 2\bar{\beta} & -
    2i\gamma_2 &  - 2i 
    \epsilon_2\\[3pt]
    (\mu+\bar{\lambda})/2 & (\lambda+\bar{\mu})/2 &
    \nu_1 & \pi_1
  \end{array}\right),
\]
The tier-1 NP equations, $\tUp^{\rho_1}{}_{ab}=0$, imply
\[\beta=-\tau/4,\quad 
\pi_1=-\tau,\quad \epsilon_2=0,\quad \Lambda= -\tau^2,\quad \lambda =
-2\gamma/3 -\bar{\mu}.\] The above constraints describe a proper
$\CH_1$ configuration.  However, the tier-2 NP equations,
$\Upsilon^{\rho_2}{}_{ab}=0$, imply
\[ \Phi_{22} = 8\gamma_1^2/9 - 2 \nu_1 \tau,\quad \gamma_2=0.\] The last
condition implies that $\fg_2=\fg_1$.  Therefore, this case does not
admit a proper $\CH_2$ configuration.

\paragraph{Case 4.} The curvature tensor has the form below.
The Petrov type is D or O, with a non-null Maxwell field.
\begin{eqnarray*}
\Psi_0=\Psi_1=\Psi_3=\Psi_4=\Phi_{00} = \Phi_{01} =
\Phi_{02}=\Phi_{12}= \Phi_{22} = 0.
\end{eqnarray*}
The automorphism group of the curvature tensor is 2-dimensional,
generated by boosts and spins $\bA_5, \bA_6$, where
\[ (\bA_\alpha) = ( \be^{14}, \be^{24}, \be^{13}, \be^{23}, \{
\be^{34}, \be^{12}\}).\] The order of the last 2 generators varies
according to the 2 subcases below.  The tier-1 connection constants
are shown below
\begin{equation*}
  (\tGu^{\rho_1}{}_a) =\left(\begin{array}{cccc}
      -\sigma & -\rho & - \tau & -\kappa \\
      -\bar{\rho} & -\bar{\sigma} & - \bar{\tau} & -\bar{\kappa} \\
      \bar{\lambda}& \bar{\mu} & \bar{\nu} & \bar{\pi} \\
      \mu & \lambda & \nu & \pi
   \end{array}\right)
\end{equation*}
A proper $\CH_2$ configuration requires $N_1=1, N_2=0$. There are 2
subcases.  Our analysis shows that neither subcase admits a proper
configuration.

\paragraph{Case 4.1.}  $G_1$ is generated by spins; $\bA_6=\be^{12}$. This
requires
\[ \kappa=\sigma=\lambda=\nu=\tau=\pi=0,\quad (\rho,\mu)\neq (0,0).\]
The Bianchi identities imply 
\[ \Psi_2 = -2 \Phi_{11}/3,\quad 
\mu_1=0,\quad \rho_1=0.\]  
From there, the tier-1 NP equations imply
\begin{eqnarray*}
  \tUp^1{}_{24}=-i\rho_2(2\epsilon_1+i\rho_2) = 0,\\
  \tUp^3{}_{23}=-i\mu_2(-2\gamma_1+i\mu_2) = 0.
\end{eqnarray*}
Hence, $\rho=\mu=0$, a contradiction.

\paragraph{Case 4.2.} $G_1$ is generated by boosts; $\bA_6 = \be^{34}$.
This requires
\[\kappa=\beta=\lambda=\nu=\rho=\mu=0,\quad
(\tau,\pi)\neq (0,0).\]
The Bianchi identities imply
\[\Psi_2=2\Phi_{11}/3,\quad \pi=\bar{\tau}.\]
From there, the tier-1 NP equations imply
\begin{eqnarray*}
  \tUp^1{}_{13} =\tau(\tau-i(\bar{\alpha}-\beta))=0\\
  \tUp^3{}_{14}=\tau(\tau+i(\bar{\alpha}+\beta))=0.
\end{eqnarray*}
Hence, $\pi=\tau = 0$, a contradiction.  

\paragraph{Case 5.}
The curvature tensor is null radiation/vacuum with an
aligned type N or conformally flat background:
\begin{eqnarray*}
  \fl \hskip 2em \Psi_0=\Psi_1=\Psi_2=\Psi_3=\Phi_{00} = \Phi_{01} =
  \Phi_{11} = 
 \Phi_{02} = \Phi_{12} = 0,\quad (\Phi_{22},\Psi_4)\neq (0,0).
\end{eqnarray*}
The automorphism group $G_0$ is 2-dimensional, generated by
null rotations $\bA_5, \bA_6$, where
\begin{equation}
  \label{eq:case5basis}
   (\bA_\alpha) = (\be^{14}, \be^{24}, \be^{34}, \be^{12},
   \be^{13}+\be^{23}, \be^{13} - \be^{23}).  
\end{equation}
The tier-1 connection constants are
\[ (\tG^{\rho_1}{}_a) = \left(\begin{array}{cccc}
    -\sigma & -\rho & - \tau & -\kappa \\
    -\bar{\rho} & -\bar{\sigma} & - \bar{\tau} & -\bar{\kappa} \\
    -\beta-\bar{\alpha} & - \alpha-\bar{\beta} & -2\gamma_1 & - 2\epsilon_1\\
    -\beta+\bar{\alpha} & - \alpha+\bar{\beta} & -2i\gamma_2 & - 2i\epsilon_2
   \end{array}\right)
\]
To obtain a proper $\CH_2$ configuration, we require $N_0=2, N_1=1$ and
$N_2=0$.  Conjugating by a spin, as necessary (the normalizer of
$G_0$ is the 4-dimensional group consisting of spins, boosts and null
rotations about $\ell$), without loss of generality we assume that $G_1$ is
generated by imaginary null rotations $\bA_6$.  This requires
\[\kappa=\sigma=\rho=\epsilon=0,\quad \alpha=\beta+\tau\neq 0. \]
The tier-2 connection constants are
\[(\tGu^{\rho_2}{}_a)= \left( \begin{array}{cccc}
    \frac{1}{2}(\mu+\bar{\lambda})  & \frac{1}{2}(\lambda+\bar{\mu}) &
    \nu_1 & \pi_1 
  \end{array}\right).\]
The tier-1 NP constraints,
\[ \tUp^1{}_{13}= \tUp^1{}_{23}=\tUp^2{}_{13} =
\tUp^2{}_{23} =  
\tUp^3{}_{12}= \tUp^4{}_{12} = \tUp^3{}_{34}=0 \]
imply
\[ \tau_2=0,\quad \beta_2=0,\quad \Lambda = -\tau^2,\quad
\pi_1=-\tau_1.\]
The NP constraint
\[ 3\,\tUp^3{}_{13} - 3\, \tUp^3{}_{3,2} - 2\,\tUp^4{}_{13} = -4 i(4\gamma_2
+3\lambda_2-3\mu_2 )(\beta+\tau)=0 \]
then implies
\[ \lambda_2 = \mu_2-4\gamma_2/3.\] The NP constraint
\begin{equation}
  \label{eq:Up413}
  \tUp^4{}_{13} = 2i\gamma_2(2\beta-\tau)=0,  
\end{equation}
means that we have to analyze two subcases: the
singular case $\gamma_2=0$, and the generic case $\gamma_2\neq 0$.

\paragraph{Case 5.1.}  Suppose that $\gamma_2=0$.  The connection
constants $\gamma_1, \tau, \beta$ are left invariant by imaginary null
rotations about $\ell$, and hence $\fg_1$ leaves invariant
$\tG^{(1)}$.  This implies that $\fg_2=\fg_1$, and therefore, the
present subcase does not admit a proper $\CH_2$ configuration.

\paragraph{Case 5.2.}
Suppose that $\gamma_2\neq 0$.
The constraint \eqref{eq:Up413} implies
\[ \beta = \tau/2.\] The assumption $\gamma_2\neq 0$ implies $\fg_2=\{
0\}$.  The field variables are
\[ (\Gamma^\lambda{}_a) = \left(
  \begin{array}{cccc}
    \frac{1}{2}(-\mu+\bar{\lambda})  & \frac{1}{2}(-\lambda+\bar{\mu}) &
    -i\nu_2 & -i\pi_2
  \end{array}\right).\]
The NP constraint
\[ \tUp^3{}_{13} = (2\gamma_1 - 3 \lambda_1-3\mu_1)\tau = 0,\]
implies that
\[ \lambda_1 = 2\gamma_1/3 - \mu_1.\]
Finally, we apply the Bianchi identities:
\begin{eqnarray*}
  \tXi^5{}_{123} = \frac{1}{2}\tau(\Psi_4-\bar{\Psi}_4) = 0\\
  \tXi^6{}_{123} = \frac{1}{2}\tau(6\Phi_{22} -\Psi_4-\bar{\Psi}_4)=0
\end{eqnarray*}
These imply
\[ \Psi_4 = 3\Phi_{22}.\]

The normalizer of the chain $G_1\subset G_0 \subset G_{-1}$ is the
3-dimensional group generated by boosts and null rotations about
$\vell$.  Conjugation by a real null rotation about $\vell$ transforms
the remaining tier-1 constants according to
\[ \tau \mapsto \tau,\quad  \gamma_1\mapsto
\gamma_1+2x(\beta_1+\tau_1),\quad \gamma_2\mapsto \gamma_2,\] where
$x$ is the constant, real-valued transformation parameter.  
A boost conjugation has the following effect:
\[ \tau\mapsto\tau, \quad \gamma\mapsto a\gamma,\] where $a\neq 0$ is
the constant, real-valued boost parameter.  Hence, conjugating by a
real null rotation and a boost, as necessary, 
 without loss
of generality we set 
\begin{equation}
  \gamma=3i/2.
\end{equation}
An imaginary null rotation transforms the remaining tier-2 constants
according to
\[ \nu_1 \mapsto \nu_1 -10x\gamma_2/3 \]
where $x$ is the constant, real-valued transformation parameter.
Hence, without loss of generality
\[ \nu_1=0.\]
The tier-2 NP constraint
$\Upsilon^5{}_{34} = 0$
implies
\[ \pi_2 = 0.\] The NP constraint $\Upsilon^5{}_{13} =0$ implies
\[ \Phi_{22} = 5\tmu_2/2-4,\quad \mu=i\tmu_2\]
where $\tmu_2$ is a real constant.

\subsection{Involutivity of the $\CH_2$ configuration}
Let $\ttau_1\neq 0, \tmu_2\neq 8/5$ be real constants.  As per the
above derivation, up to $\O(\eta)$ equivalence, the most general proper
$\CH_2$ configuration is given by
\begin{eqnarray}
  &&\tau=-\pi=2\beta=2\alpha/3=\ttau_1,\\
  &&\gamma= 3i/2,\\
  &&\mu=\lambda+2i = i \tmu_2,\\
  &&\nu_1=0,\\
  &&\Phi_{22}=\Psi_4/3 = -4 + 5 \tmu_2/2,\\
  &&\Lambda=-\ttau_1^2,
\end{eqnarray}
with all other connection and curvature scalars equal to zero.  
Equivalently, relative to
the basis \eqref{eq:case5basis}, the above configuration may be
described as follows:
\begin{eqnarray}
  \label{eq:bGamma1def}
  &&\bGamma^1 = \bGamma^2 = - \ttau_1\bom^3,\\
  &&\bGamma^3 = -2\ttau_1\, (\bom^1+\bom^2),\\
  &&\bGamma^4 = \ttau_1\, (\bom^1-\bom^2) - 3\rmi\, \bom^3,\\
  &&\bGamma^5 = \rmi\,(\bom^1-\bom^2) + \rmi \ttau_1\,, \bom^4\\
  &&\bGamma^6 = \rmi(1-\tmu_2)(\bom^1+\bom^2)-\rmi\nu_2\, \bom^3,\\
  &&\bOmega^1 = 2\ttau_1^2\, \bom^2\wedge \bom^3,\\
  &&\bOmega^2 = 2\ttau_1^2\, \bom^1\wedge \bom^3,\\
  &&\bOmega^3 = 2\ttau_1^2\, \bom^3\wedge \bom^4,\\
  &&\bOmega^4 = 2\ttau_1^2\, \bom^1\wedge \bom^2,\\
  &&\bOmega^5 = (5\tmu_2-8)(\bom^1+\bom^2)\wedge \bom^3 + \ttau_1^2\,
  (\bom^1+\bom^2)\wedge \bom^4,\\
  \label{eq:bOmega6def}
  &&\bOmega^6 = ((5/2)\tmu_2-4)(\bom^1-\bom^2)\wedge \bom^3 - \ttau_1^2\,
  (\bom^1-\bom^2)\wedge \bom^4.
\end{eqnarray}
The EDS  for this configuration corresponds to the following structure
equations:
\begin{eqnarray}
  &&\rmd \bom^1  = -\bom^1\wedge\bGamma^4+\bom^3\wedge(\bGamma^5-\bGamma^6) +
  \bom^4\wedge \bGamma^2, \\
  &&\rmd \bom^2  = \bom^2\wedge\bGamma^4+\bom^3\wedge(\bGamma^5+\bGamma^6) +
  \bom^4\wedge \bGamma^1, \\
  &&\rmd\bom^3 = \bom^1\wedge\bGamma^1
  +\bom^2\wedge\bGamma^2+\bom^3\wedge\bGamma^3 ,\\
  &&\rmd\bom^4 = \bom^1\wedge(\bGamma^5+\bGamma^6) + \bom^2\wedge
  (\bGamma^5-\bGamma^6) -\bom^4\wedge\bGamma^3,\\
  &&\rmd\bGamma^6= \bGamma^3\wedge\bGamma^4-\bGamma^4\wedge\bGamma^5 +
  \bOmega^6
\end{eqnarray}
Note that the structure equations 
\[\rmd\bGamma^\alpha=\ldots,\qquad
\alpha=1,\ldots, 5\]
 are satisfied identically, by construction.
Substituting \eqref{eq:bGamma1def}-\eqref{eq:bOmega6def} into the
above gives
{\small
\begin{eqnarray}
  \label{eq:ch2dbom1}
 &&\rmd \bom^1 =\ttau_1\,\bom^1\wedge\bom^2 - \rmi
 (\tmu_2-3)\,\bom^1\wedge 
  \bom^3 - \rmi (\tmu_2-2)\, \bom^2\wedge\bom^3,\\
  &&\rmd \bom^2=-\ttau_1\,\bom^1\wedge\bom^2+ \rmi (\tmu_2-2)\,
  \bom^1\wedge\bom^3  + \rmi (\tmu_2-3)\,\bom^2\wedge \bom^3 ,\\
  &&\rmd \bom^3 =\ttau_1\, (\bom^1+\bom^2)\wedge \bom^3,\\
  \label{eq:ch2dbom4}
  &&\rmd \bom^4 =- 2\rmi\tmu_2\, \bom^1\wedge\bom^2 -\rmi\nu_2
  (\bom^1-\bom^2)\wedge \bom^3 - 3\ttau_1\,
  (\bom^1+\bom^2)\wedge\bom^4,\\
  \label{eq:ch2npred}
  &&\rmd \nu_2\wedge\bom^3= \big( (3\rmi/2)\,\tmu_2\,
  (\bom^1-\bom^2) - 3\ttau_1\nu_2 \,
  (\bom^1+\bom^2)-3\ttau_1 \bom^{4}\big)\wedge \bom^3.
\end{eqnarray}}
The scalar form of equations \eqref{eq:ch2dbom1}-\eqref{eq:ch2dbom4}
is:
{\small
\begin{eqnarray}
  \label{eq:ch2y1}
  &&\fl \quad y^1{}_{[i,j]} = -\ttau_1\, y^1{}_{[i} y^2{}_{j]} + \rmi
  (\tmu_2-3)\,y^1{}_{[i} y^3{}_{j]}
  +\rmi(\tmu_2-2)\,y^1{}_{[i}y^3{}_{j]},\\
  &&\fl \quad y^2{}_{[i,j]} =  \ttau_1\, y^1{}_{[i} y^2{}_{j]} - \rmi
  (\tmu_2-2)\,y^1{}_{[i} y^3{}_{j]}
  -\rmi(\tmu_2-3)\,y^1{}_{[i}y^3{}_{j]},\\
  &&\fl \quad y^3{}_{[i,j]} = -\ttau_1 (y^1{}_{[i} y^3{j]} +
  y^2{}_{[i} y^3{}_{j]}),\\ 
  \label{eq:ch2y4}
  &&\fl\quad y^4{}_{[i,j]} = 2\rmi\tmu_2y^1{}_{[i} y^2{}_{j]}
  +3\ttau_1\left(y^1{}_{[i}y^4{}_{j]} +y^2{}_{[i}y^4{}_{j]} \right) +2\nu_2\rmi\left(y^1{}_{[i}y^3{}_{j]}
    -y^2{}_{[i}y^3{}_{j]}\right),
\end{eqnarray}}
The scalar form of the constrained NP equations \eqref{eq:ch2npred} is
\begin{eqnarray}
  \label{eq:ch2Dnu}
  &&D\nu_2 = -3\ttau_1,\\
  &&\delta\nu_2 = -3\ttau_1\nu_2+(3\rmi/2)\,\tmu_2,\\
  \label{eq:ch2delstarnu}
  &&\delta^*\nu_2 = -3\ttau_1\nu_2-(3\rmi/2)\,\tmu_2,
\end{eqnarray}
where as per \eqref{eq:yaidef}, the derivative of  a scalar $f$ is
given by
\begin{equation}
  f_{,i} = y^1{}_i \delta f + y^2{}_i \delta^* f +
  y^3{}_i \Delta f + y^4{}_i D f.
\end{equation}
The scalar Bianchi equations are identically satisfied, by
construction.

By inspection, equations \eqref{eq:ch2y1}-\eqref{eq:ch2delstarnu} have
no zeroth order integrability conditions; it isn't possible to
eliminate all the derivatives from these equations.  
To prove involutivity, we rewrite
\eqref{eq:ch2dbom1}-\eqref{eq:ch2dbom4} as commutator relations:
\begin{eqnarray}
  \label{eq:ch2comm1}
 && \delta^* \delta-\delta\delta^* =  -2\rmi\, \tmu_2 D+\ttau_1\,
 (\delta-\delta^*) ,\\
 && \Delta\delta -\delta\Delta =\rmi(3-\tmu_2)\delta
 +\rmi(\tmu_2-2)\delta^* +\ttau_1\Delta -\rmi\nu_2\, D,\\
 &&D\delta-\delta D =  -3\ttau_1 D,\\
  \label{eq:ch2comm6}
 &&D\Delta-\Delta D = 0.
\end{eqnarray}
Combining these with \eqref{eq:ch2Dnu}-\eqref{eq:ch2delstarnu} yields
the following first-order relations:
\begin{eqnarray}
 && 3\ttau_1\big((\delta-\delta^*)\nu_2-3\rmi\, \tmu_2\big)=0,\\
 && 3\ttau_1\big(D\nu_2 +3\ttau_1\big) = 0.
\end{eqnarray}
Since no independent first-order relations are implied, the above
system of equations is involutive.

\subsection{The $\CH_2$ exact solution}
Next, we integrate the structure equations
\eqref{eq:ch2dbom1}-\eqref{eq:ch2dbom4} and describe the most general
proper $\CH_2$ spacetime as an exact solution.
At this point, it is convenient to introduce the 1-forms
\begin{eqnarray}
  \label{eq:tth1def}
  &&\bth^1 = \ttau_1/2\, (\bom^1+\bom^2),\\
  &&\bth^2 = -i\ttau_1/2\, (\bom^1-\bom^2)+(2\tmu_2-5)/4\,
  \bom^3,\\
  &&\bth^3 = \bom^3,\\
  \label{eq:tth4def}
  &&\bth^4 = (\tmu_2/\ttau_1)(-i/2 \, (\bom^1-\bom^2)
  +(2\tmu_2-5)/4\, \bom^3) + \bom^4,\\
  \label{eq:tth5def}
  &&\bth^5= ((1-\tmu_2)(\bom^1+\bom^2)
  + i \bGamma^6)/\ttau_1 = (\nu_2/\ttau_1)\, \bom^3 .
\end{eqnarray}
The structure equations now assume a particularly simple form:
\begin{eqnarray}
  &&\rmd\bth^1 = -\bth^2\wedge \bth^3,\\
  &&\rmd\bth^2 = -2\,\bth^1\wedge \bth^2,\\
  &&\rmd\bth^3 = 2\,\bth^1\wedge \bth^3,\\
  \label{eq:tth4seq}
  &&\rmd\bth^4 = -6\,\bth^1\wedge\bth^4 + 2\,\bth^2\wedge\bth^5,\\
  \label{eq:tth5seq}
  &&\rmd\bth^5 = -4\,\bth^1\wedge\bth^5 +3\,\bth^3\wedge\bth^4.
\end{eqnarray}
The first 3 equations are the structure equations for $\SL_2$.  Hence,
$\bth^1,\bth^2,\bth^3$ can be integrated by means of local coordinates
on $\SL_2$.    We choose the coordinatization
\[
\left( \begin{array}{cc}
    1&0\\s&1
  \end{array}\right)
\left( \begin{array}{cc}
    e^{b/2}&0\\0&e^{-b/2}
  \end{array}\right)
\left( \begin{array}{cc}
    1&a\\0&1
  \end{array}\right)
.
\]
This yields the following expressions for the Maurer-Cartan forms:
\begin{eqnarray}
  \label{eq:tth1form}
  &&\bth^1 = \rmd b/2-ae^b \rmd s,\\
  &&\bth^2 = \rmd a+ a \rmd b- a^2 e^b \rmd s,\\
  \label{eq:tth3form}
  &&\bth^3 = e^b \, \rmd s
\end{eqnarray}
Substituting \eqref{eq:tth5def} into \eqref{eq:tth5seq} we obtain
\begin{equation}
  \bth^4 \equiv -(\nu_2 \rmd b +\rmd\nu_2)/\ttau_1\quad \textrm{mod}\; \rmd s.
\end{equation}
Writing
\begin{equation}
  \label{eq:tth4form}
  \bth^4 =  -(\nu_2 \rmd b +\rmd\nu_2)/\ttau_1+ (2ae^b\nu_2 + e^{-3b} F) \rmd s,
\end{equation}
and substituting into \eqref{eq:tth4seq} gives
\[ e^{-3b} \rmd F \wedge \rmd s = 0.\] Hence $F=F(s)$.  Up to a choice of
local coordinates and a choice of the function $F(s)$, the above
solution is the most general possible.  Indeed, we could take
$a,b,s,\nu_2$ as coordinates. However, it will be more convenient to
set
\begin{equation}
  \nu_2 = -3\ttau_1 e^{-3b} t,
\end{equation}
and to use $a,b,s,t$ as coordinates.  Finally, substituting
\eqref{eq:tth1form}-\eqref{eq:tth3form}, \eqref{eq:tth4form} into
\eqref{eq:tth1def}-\eqref{eq:tth4def}, we obtain the form of the
general (up to a change of coordinates) solution of the $\CH_2$ field
equations:
\begin{eqnarray}
  \label{eq:ch2om1}
  \bom^1 &=& \left(\rmd b/2 + \rmi(a\, \rmd b+ \rmd a)-e^b \left( a +
      \rmi\left(a^2+\tmu_2/2-5/4\right) \right) \rmd s
  \right)/\ttau_1,\\       
  \bom^2 &=& \left(\rmd b/2 - \rmi(a\, \rmd b+ \rmd a)-e^b \left( a -
      \rmi\left(a^2+\tmu_2/2-5/4\right) \right) \rmd s
  \right)/\ttau_1,\\       
  \bom^3 &=& e^b \rmd s,\\
  \label{eq:ch2om4}
  \bom^4 &=& e^{-3b}\rmd t- (\tmu_2/\ttau_1^2)(\rmd a+a\, \rmd b) \\ \nonumber
  &&+\left(F(s)e^{-3b}- 6\,
    ae^{-2b}t 
    +(\tmu_2/\ttau_1^2) \left(a^2+\tmu_2/4-5/8\right)e^b \right)\rmd s.
\end{eqnarray}
\subsection{The $\CH_2$ equivalence problem and Killing vectors}
Next, we solve the local equivalence problem for the class of proper
$\CH_2$ spacetimes, as described by \eqref{eq:ch2om1}
-\eqref{eq:ch2om4}.  In our analysis, we use ideas from Chapter 8 of
Olver\cite{olver}, as well as Karlhede's algorithm.  Again, it will be
more convenient to work with the coframe $\bth^a$ defined in
\eqref{eq:tth1def} - \eqref{eq:tth4def}.  The latter differs from the
canonical null-orthogonal tetrad $\bom^a$ by a constant linear
transformation, so both coframes will yield the same differential
invariants.

The structure functions in  equations
\eqref{eq:ch2dbom1}-\eqref{eq:ch2dbom4} yield 
our first differential invariants, namely the constants
$\ttau_1,\tmu_1$ and the scalar $\nu_2$.  The constrained NP equations
\eqref{eq:ch2Dnu} -\eqref{eq:ch2delstarnu}  indicate that $\delta \nu_2,
\delta^*\nu_2, \Delta \nu_2$  are all functionally dependent on
$\nu_2$.  Thus the only candidate for an independent differential
invariant is $D\nu_2$.  The commutator
relations \eqref{eq:ch2comm1}-\eqref{eq:ch2comm6} show that $\delta
D\nu_2, \delta^* D\nu_2, \Delta D\nu_2$ are all functionally
dependent on $\nu_2, D\nu_2$.  Proceeding inductively, we have 
that 
\begin{equation}
  \nu_2,\; D\nu_2,\;  D^2 \nu_2,\;  D^3\nu_2
\end{equation}
is a maximal set of functionally independent differential invariants.

However, since we have the exact solution \eqref{eq:ch2om1} -
\eqref{eq:ch2om4} the analysis of the equivalence problem can be
considerably simplified.  Let us express the first differential
invariant as
\[ I_1 := \nu_2 /(3\ttau_1) = e^{-3b} t.\] Since
\[ R^3 = \Gamma^{(3)}\cdot \tR^2,\] this differential invariant arises
as a component of $R^3$ taken relative to the preferred tetrad.
Working relative to the preferred tetrad, we have
\[ \rmd I_1 = -6 I_1 \bth^1- e^{-4b}F(s) \bth^3 + \bth^4.\] There are 3
cases to consider.

\paragraph{Case 5.2.1.}  Suppose that $F(s)\neq 0$.  
Let us also set
\begin{eqnarray}
  \label{eq:F1def}
  &&F_1(s) = F'(s)/F(s),\\
  \label{eq:F2def}
  &&F_2(s) = (F_1'(s) -F_1^2(s)/8)/\sqrt{|F(s)|}.  
\end{eqnarray}
We now have a
second functionally independent differential invariant, namely
\[ I_2 := \log|F| - 4b.\]
Since
\[ R^4 = \rmd R^3 + \Gamma^{(3)}\cdot R^3,\] this is the only
functionally independent invariant arising from $R^4$.  We have
\[ \rmd I_2 = -8 \bth^1 +I_3\, \bth^3,\] where 
\begin{equation}
  \label{eq:I3def}
   I_3 := e^{-b} F_1(s)-8a.
\end{equation}
Since
\[ R^5 = \rmd R^4 + \Gamma^{(3)}\cdot R^4,\] this is the only
functionally independent invariant arising from $R^5$.  Continuing, 
\[ 
 \rmd I_3 = -2I_3 \bth^1 -8 \bth^2 +\left(I_3^2/8 + e^{I_2/2}
  I_4\right) \bth^3,\]
where
\begin{equation}
  \label{eq:I4def}
  I_4 := F_2(s).  
\end{equation}
Continuing,
\[ \rmd I_4 = F_2'(s)\, \rmd s = e^{-b} F_2'(s) \bth^3 = e^{I_2/4}I_5
\bth^3,\]
where
\begin{equation}
  \label{eq:I5def}
  I_5:= F_2'(s)|F(s)|^{-1/4}
\end{equation}
is a differential invariant arising from $R^7$.

Now there are two subcases.  Generically $F_2'(s)\neq 0$, and hence
$I_4$ is a functionally independent invariant, arising from $R^6$.
Since both $I_4$ and $I_5$ are functions of $s$, locally
\[ I_5 = \phi(I_4).\] Therefore, the classification problem is solved
by means of the essential constants $\ttau_1, \tmu_2$ and an essential
parameter function $\phi(x)$.  Therefore, generically, $q_M=7$; the IC
of our spacetime requires $R^7$.

\noindent
\paragraph{Case 5.2.2.} Suppose that $I_4(s)$ is a constant.  In this
case, $q_M=6$.  The essential constants $\ttau_1,\tmu_2,I_4$ solve the
IC problem.  Since there are only 3 functionally independent
invariants. The Lie algebra of Killing vectors is 1-dimensional.

\paragraph{Case 5.2.3.} If $F(s) = 0$, then the preferred tetrad
possesses only one functionally independent invariant.  Hence, the
spacetime has a 3-dimensional isometry group, which is isomorphic to
$\SL_2\Rset$. The orbits are given by $\nu_2 = \mbox{const.}$ In this
case, $q_M=4$.

\subsection{The proper $\CH_2$ spacetimes}
The above solution represents a spacetime where coupled gravity and
electromagnetic waves propagate in a negatively curved background.
The above proper $\CH_2$ metric belongs to a general family of such
spacetimes, first described in \cite{GP81,orr}.  However, up to now it
was not known that these solutions contained a $\CH_2$ subfamily.  Let
$\tLambda<0$ be a negative constant.  Following \cite{orr}, the
general exact solution has the form
\begin{equation}
  \fl g_{ij} dx^i dx^j= 2p^{-2} d\zeta d\bar{\zeta} - 2 q^2 p^{-2}\,
  (( -(\tLambda A^2 + B\bar{B})r^2+ r\, q_s/q +
  2 H p/q) 
  \rmd s + d r)d s,
\end{equation}
where $\zeta,\bar{\zeta},r,s$ are coordinates, and where
\begin{eqnarray}
  &&p= 1+\tLambda \zeta \bar{\zeta},\\
  &&q = (1-\tLambda \zeta \bar{\zeta}) A +\bar{B}\zeta+B\bar{\zeta},\quad
  A=\bar{A},\; A=A(s),\; B=B(s)\\
  &&H_{\zeta\bar{\zeta}} + 2 \tLambda p^{-2} H = f\bar{f} p/q,\quad f=f(\zeta,s),
\end{eqnarray}
and where
\begin{equation}
  f \, \rmd\zeta\wedge \rmd s +\bar{f}\, \rmd\bar{\zeta}\wedge \rmd s 
\end{equation}
is the electromagnetic field.  The proper $\CH_2$ solution described
above is a particular subclass of such spacetimes.  In order to obtain
the $\CH_2$ specialization, one has to change coordinates and
specialize the parameters of the general ansatz as follows:
\begin{eqnarray}
  &&\tLambda=-\ttau_1^2,\quad   A = 1,\quad B= -e^{3is} \ttau_1,\\ 
  &&H =[36-72/p+(27+16\ttau_1^2 F(s)) q/p + 
  (10\tmu_2-16)p^3/q^3]/(32\ttau_1^2) 
  \\ 
  && a = (\ttau_1/p) \,\Im(e^{-3is} \zeta),\\
  && b =
  \log(p)-\log(q)\\ 
  &&t= r+a e^b(3/2+e^{2b}(1+4a^2/3))/\ttau_1^2.
\end{eqnarray}
As was mentioned above, if $F(s)=0$, then the metric possesses an
$\SL_2\Rset$ isometry group and $q_M=4$.  However, for generic $F(s)$,
the specialized metric admits no Killing vectors and has IC order
$q_M=7$.

\section{Conclusion}
We have analyzed 4-dimensional Lorentzian curvature homogeneous
manifolds in terms of an exterior differential system and have proved
that $k_{1,3}=3$.  Therefore for any 4-dimensional, Lorentzian $M$ the
CH$_3$ conditions imply that $M$ is locally homogeneous.  In addition,
the class of proper CH$_2$ geometries has been explicitly determined
in equations (\ref{eq:ch2om1})--(\ref{eq:ch2om4}), these provide a
counterexample to a conjecture of Gilkey stating that $k_{1,3}=2$.

In regards to the invariant classification problem, it has been shown
that, generically, (\ref{eq:ch2om1})--(\ref{eq:ch2om4}) have IC order
$q_M=7$, thereby settling a long-standing question about the Karlhede
bound for four-dimensional, Lorentz-signature spacetimes.  The
curvature tensor along with its first and second covariant derivatives
completely fix the frame and provide two essential constants:
$\ttau_1, \tmu_2$.  Generically, the higher order covariant
derivatives of the curvature tensor give rise to five differential
invariants $I_1,\ldots, I_4, I_5=\phi(I_4)$.  The first 4 are
functionally independent.  The essential parameters $\ttau_1, \tmu_2,
\phi(x)$ invariantly classify the spacetime.  There are 2 singular
subfamilies.  The parameter function $F(s)$ is not, by itself, an
invariant.  However, the condition $F(s)=0$ is invariant. The
corresponding spacetime has 3 Killing vectors.  The subfamily
characterized by the condition $F_2'(s)=0$ has 1 Killing vector.

The above $\CH_2$ solutions have constant zero-order
curvature invariants and vanishing higher order curvature
invariants\footnote{A thorough study of the curvature invariants of
  vacuum P-type N with $\Lambda \neq 0$ is presented in
  \cite{pravda98}.}.  Thus, these solutions are examples of constant
scalar invariant spacetimes (CSI) \cite{csi}.  However, it may be more
natural to regard them as vanishing scalar invariant spacetimes (VSI)
\cite{vsi4d} \cite{vsiNd} with a cosmological constant since only
zeroth order invariants are nonzero constants.  This indicates a
slight modification required in the CSI$_{R}$ conjecture \cite{csi}
where we extend VSI to also include a cosmological constant.  It is
curious that the curvature invariants cannot be used to solve the
invariant classification problem.  This was also shown to occur in
Einstein solvmanifolds \cite{hervik04}.

The proper $\CH_2$ solutions describe gravitational waves and
electromagnetic radiation propagating in an anti-de Sitter background.
These are contained in the class of metrics presented in \cite{orr}
(see also \cite{GP81}).  Generalizations and further analysis was
subsequently given in \cite{bicak991} \cite{bicak992} \cite{podo03}
\cite{griffiths04} and extended to higher dimensions in
\cite{obukhov04}.  This class has been applicable in a number of
important areas in the literature, such as in the study of
Einstein-Yang-Mills solutions \cite{guven79} which were more recently
investigated to determine P-type III solutions \cite{fuster05}.  In
addition, they arise in Lovelock-Yang-Mills theory \cite{gleiser05},
in the theory of metric-affine gravity \cite{obukhov06}
\cite{garcia00} and in supergravity \cite{cariglia04} \cite{kerimo05}.
A consideration of the proper CH$_2$ class within this context may
give solutions with interesting properties, or at the least may have
practical relevance since the components of the curvature tensor up to
its second covariant derivative is constant.

It is quite remarkable that the $q_M=7$ condition precisely picks out
this one particular family of spacetimes with its particular physics.
Note that if $\tmu_2=8/5$, we obtain anti-de Sitter spacetime, and the
choice of $F(s)$ becomes irrelevant.  However, generically the nature
of $F(s)$ must have some phenomenological interpretation in terms of
the gravity and electromagnetic radiation, albeit one that requires
seventh order information.  Why does this not occur for flat
$\Lambda=0$ space or for deSitter $\Lambda>0$ space?

In this paper we have also illustrated the applicability of exterior
differential systems to the study of the curvature homogeneity
problem.  EDS has applications in the study of the Weyl-Lanczos
problem \cite{DoGe04} and in the analysis of vacuum solutions
\cite{Estabrook06}.  It is natural to expect that the use of EDS in
the study of exact solutions could provide some further insights in
relativity.

\section*{Acknowledgments}T
The research of RM is supported in part by NSERC grant
RGPIN-228057-2004.  We thank J. \AA man for useful comments.

\end{document}